\documentclass[aps,prl,twocolumn,showpacs,10pt,floatfix,superscriptaddress,floatfix, longbibliography]{revtex4-1}
\usepackage{amsmath}
\usepackage{braket}
\usepackage{amssymb}
\usepackage[normalem]{ulem}
\usepackage{ dsfont }
\usepackage{bm}
\usepackage[dvipsnames]{xcolor}
\usepackage{stmaryrd}
\usepackage{bbm}
\usepackage{mathtools}

\newcommand{\vk}{{\vec k}}

\newcommand{\floor}[1]{\lfloor #1 \rfloor}

\newcommand{\vR}{\vec R}
\renewcommand{\vr}{\vec r}

\newcommand{\vx}{\vec x}

\newcommand{\vv}{\vec v}

\begin{document}

\title{Topologically localized insulators}
\author{Bastien Lapierre, Titus Neupert, Luka Trifunovic}
\affiliation{Department of Physics, University of Zurich, Winterthurerstrasse 190, 8057 Zurich, Switzerland}
\date{\today}
\begin{abstract}
We show that fully-localized, three-dimensional, time-reversal-symmetry-broken insulators do not belong to a single phase of matter but can realize topologically distinct phases that are labelled by integers. The phase transition occurs only when the system becomes conducting at some filling. We find that these novel topological phases are fundamentally distinct from insulators without disorder: they are guaranteed to host delocalized boundary states giving rise to the quantized boundary Hall conductance, whose value is equal to the bulk topological invariant.
\end{abstract}
\maketitle   

\textit{Introduction} ---
The discovery of quantum Hall effect revealed that topology, a branch of
mathematics, plays an important role for understanding properties of quantum
materials. Topological quantum materials are typically bulk insulators with
boundaries that are perfect metals~\cite{ryu2007,schnyder2008}. Further
developments~\cite{kitaev2001,schnyder2008} showed that superconductors can be
topological, too. These developments led to the complete classification of
(non-interacting) topological phases, the result known as tenfold-way or ``periodic
table of topological phases of matter''~\cite{kitaev2009,schnyder2009}. The
modes that appear on the boundary of topological quantum materials are called
topological and anomalous --- they typically give rise to either quantized electrical or
thermal responses and promise many applications in the areas of quantum
computing~\cite{kitaev2001}, backscattering-free quantum transport, and even
catalysis~\cite{chen2011}.

More
recently~\cite{schindler2018,song2017,langbehn2017,trifunovic2020,papenbrock2002, PhysRevLett.125.266804, PhysRevB.104.134508, PhysRevB.103.115308, PhysRevB.103.085408},
it was shown that crystalline symmetry adds a new twist to the topological
classification: a three-dimensional topological crystalline insulator or superconductor
can have topological modes (states) appearing on its hinges (corners) while its 
faces (and hinges) are insulating.
Interestingly~\cite{benalcazar2017,trifunovic2020,watanabe2021}, crystalline
symmetry can turn an atomic insulator topological, in which case they
may or may not have fractionally quantized boundary charges. Moreover, the
crystalline symmetry facilitates the discovery of topological materials~\cite{bradlyn2017,po2017,geier2020,Skurativska2020,Ono2019}, with
many thousand candidates being predicted~\cite{bradlyn2017,zhang2019} and some
experimentally confirmed~\cite{koenig2007,wu2018,xia2009,ran2019,Murani2017}.

Not only the boundary of a topological material but also its bulk electronic states have
intriguing properties: all topological insulators, in the absence
of crystalline and sublattice symmetries, have an obstruction to full
localization of its occupied bulk electronic states~\footnote{A similar statement in case of topological band insulators in the absence of crystalline and sublattice symmetries: there is an obstruction to spanning the Hilbert space of occupied electronic states by exponentially localized or compact Wannier functions~\cite{read2017}}. This property is best
studied in the case of quantum Hall insulators~\cite{chalker1988}, where in
the presence of disorder topology guarantees the existence of a single energy per Landau level where delocalized states appear. Similarly, the obstruction
to full localization was established for the case of quantum spin
Hall~\cite{onoda2007} and three-dimensional topological
insulators~\cite{morimoto2015}. Hence, within the tenfold-way paradigm, a fully
localized insulator (i.e., an Anderson insulator at all fillings) is guaranteed to be
topologically trivial.
\begin{figure}[t]
	\centering
	\includegraphics[width=\columnwidth]{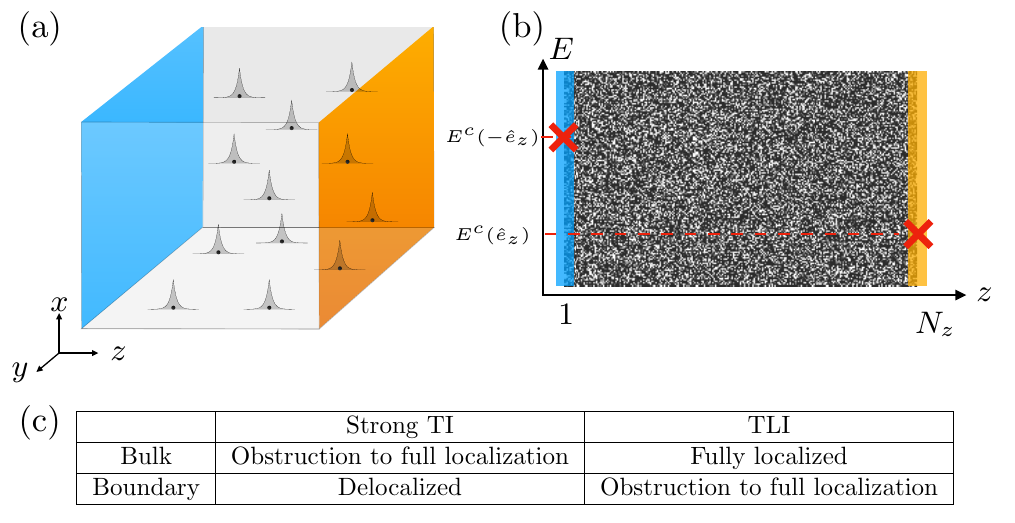}
	\caption{(a) TLI with $N_\text{TLI}=1$ for open boundary conditions along $z$-direction only: all the bulk states are localized by disorder,
		while the two boundary surfaces (orange and blue) host
		delocalized states and have opposite quantized Hall
		conductance. (b) The energies $E^\text{c}(\pm\hat e_z)$ of delocalized states are not constrained by the bulk. (c) Comparison between TLIs and strong TIs. (A TI is called strong if its existence does not rely on translational symmetry.)
		}
	\label{fig:TLI}
\end{figure}

In this work, we introduce a new notion of topology that applies to fully
localized insulators, i.e., Anderson insulator at \textit{all} fillings, in contrast to tenfold-way topological phases which are required to be insulating only at one particular filling~\footnote{Topological Anderson Insulators~\cite{li2009}, that are captured by tenfold-way classification~\cite{kitaev2001,schnyder2008}, have delocalized states in the bulk and hence do not represent fully localized insulators.}. In particular, we consider three-dimensional systems
without time-reversal symmetry and find topologically distinct
fully localized insulators that can be labelled by integers. The phase
transition can occur only if the system becomes
conducting at some filling. We refer to these topologically non-trivial phases as
\textit{topologically localized insulators} (TLIs). We show that, although all
the bulk states of a TLI are exponentially localized, there is an obstruction
to localizing them all the way down to an atomic limit. Importantly, 
electronic states with  support close to the boundary carry quantized Hall
conductance which can be measured in the Corbino geometry via flux insertion.
Note that, for the slab geometry in Fig.~\ref{fig:TLI}a, the boundary consists
of two disjoint planes with normals $\pm\hat e_z$, and the Hall conductances of
these two planes are quantized to the opposite value due to their opposite orientation. Furthermore,
at each of these planes, topologically protected delocalized
states~\cite{chalker1988} emerge and remain so under an arbitrarily strong disorder at
the boundary, see Fig.~\ref{fig:TLI}b--c. We conclude that the boundary of a TLI is anomalous since it cannot
be realized as a two-dimensional system: a disordered two-dimensional Chern
insulator can host delocalized bulk states only up to disorder strengths that do not
close its mobility gap~\cite{prodan2011}. (A Chern insulator is a locally finite-dimensional Hilbert-space version of Quantum Hall Insulator.)

\begin{figure}[t]
	\centering
	\includegraphics[width=\columnwidth]{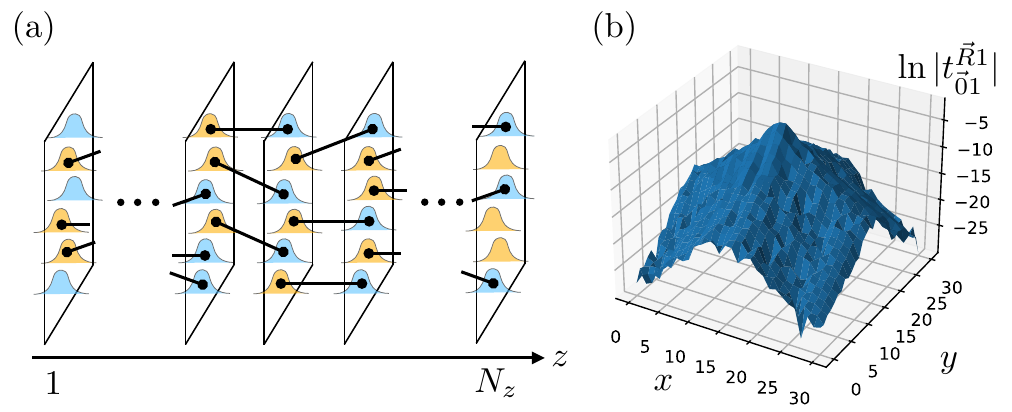}
	\caption{(a) TLI construction from stack of two-dimensional layers in $z$-direction. The Hilbert space of each layer is divided into two subspaces spanned by blue and orange orbitals: occupying only blue (orange) orbitals with electrons results in quantized Hall conductance of the layer $\sigma_{12}=e^2/h$ ($\sigma_{12}=-e^2/h$). The blue (orange) orbitals cannot all have exponential localization; Exponential localization is obtained by hybridizing differently colored orbitals from neighbouring layers. (b) Distance dependence of the hopping elements $\ln |t_{\vec{0}1}^{\vR 1}|$ of the TLI with $\vR=(x,y,0)$ for the system size $N_x=N_y=31$, where $t_{\vec{0}1}^{\vR 1}$ is defined below Eq.~(\ref{eq:Hzz1}). We observe an exponential decay of the hopping elements $t_{\vec{0}1}^{\vR 1}$ in all directions. }
\label{fig:const}
\end{figure}

\textit{Model} --- We construct a TLI by stacking two-dimensional layers in
$z$-direction. We divide the Hilbert space spanned by electronic orbitals of a
single layer into the blue and the orange subspaces, see Fig.~\ref{fig:const}a.
We require that for the blue (orange) subspace
filled with electrons, the layer has Hall conductance $\sigma_{xy}=e^2/h$
($\sigma_{xy}=-e^2/h$). Hence, not all blue (orange) orbitals in
Fig.~\ref{fig:const}(a) can be exponentially localized. On the other hand, if we
couple the blue orbitals from the layer $z$ to the orange orbitals from the
layer $z+1$, the hybridized orbitals can be all exponentially localized, as in
the case of eigenstates of the following tight-binding model of a TLI
\begin{equation}
H =\sum_{\vR\alpha,\vR^\prime\alpha^\prime}t_{\vR\alpha}^{\vR^\prime\alpha^\prime}\ket{\vec{R}\alpha}\bra{\vec{R}^\prime\alpha^\prime},\\
\label{eq:Hzz1}
\end{equation}
with the hopping amplitudes and onsite potentials (for $\vR=\vR^\prime$) expressed as
\begin{equation}
t_{\vR\alpha}^{\vR^\prime\alpha^\prime}=\sum_{\vR^{\prime\prime}\alpha^{\prime\prime}}
W_{\vec{R}^{\prime\prime}\alpha^{\prime\prime}}C_{\vR\alpha}^{\vR^{\prime\prime}\alpha^{\prime\prime}}(C_{\vR^\prime\alpha^\prime}^{\vR^{\prime\prime}\alpha^{\prime\prime}})^*.
\end{equation}
The atomic orbitals of the above stack of layers are denoted by $\ket{\vR\alpha}$,
with $\alpha\in\{1,2\}$ an orbital degree of freedom and $\vR$ a
three-dimensional cubic lattice vector. $W_{\vec{R}\alpha}$ are independent,
uniformly distributed in $[-W,W]$, random real numbers. The
coefficients $C_{\vR^\prime\alpha^\prime}^{\vR\alpha}$ are defined through the
lattice vector basis expansion of the wavefunctions
\begin{equation}
\ket{w_{\vR\alpha}}\equiv \mathcal{P}_{z}^-\ket{\vR\alpha}+\mathcal{P}_{z+1}^+\ket{(\vR+\hat{e}_z)\alpha},
\end{equation}
i.e., $\ket{w_{\vR\alpha}}=\sum_{\vR'\alpha'}C_{\vR'\alpha'}^{\vR\alpha}\ket{\vR'\alpha'}$,
where $\mathcal{P}_z^+$ ($\mathcal{P}_z^-$) is the projector onto the blue (orange) subspace of the
layer $z$. Such $\mathcal{P}_z^+$ and $\mathcal{P}_z^-$ can be obtained as the projectors on the occupied and empty bands of a two-band (disorder-free) Chern insulator defined on layer $z$ (see~\cite{SM} for a concrete model). It is crucial to note that $\ket{w_{\vR\alpha}}$  are exponentially
localized, and orthonormal $\langle
w_{\vR^\prime\alpha^\prime}\ket{w_{\vR\alpha}}=\delta_{\vR\vR^\prime}\delta_{\alpha\alpha^\prime}$,
leading to exponentially decaying matrix elements
$t_{\vR\alpha}^{\vR^\prime\alpha^\prime}$, see Fig.~\ref{fig:const}b. The
exponential localization of $\ket{w_{\vR\alpha}}$ follows from that of
$\mathcal{P}_z^\pm\ket{\vR\alpha}$~\cite{thonhauser2006}, whereas orthonormality can be
satisfied only for the Hilbert space onto which $(\mathcal{P}_{z}^-+\mathcal{P}_{z+1}^+)$ projects,
see~\cite{SM}. The states $\ket{w_{\vR\alpha}}$ form a complete set of localized
eigenstates of $H$ under periodic boundary conditions (PBC).

From the above construction it is evident that for a $z$-terminated crystal,
there are unpaired blue (orange) orbitals on the layer $z=1$ ($z=N_z$) which,
when filled with electrons, give quantized Hall conductance $\sigma_{xy}=e^2/h$
($\sigma_{xy}=-e^2/h$), see Fig.~\ref{fig:TLI}. Additionally, if a magnetic flux
quantum $\Phi_0=h/e$ is threaded through the layers, the blue subspace of each
layer expands to accommodate one more electron while the orange subspace
shrinks by the same amount --- this statement is know as Streda's
theorem~\cite{streda1982}. Hence, if the bulk is fully filled with electrons,
the applied flux $\Phi_0$ transfers one electron from layer $z$ to layer
$z+1$ resulting in the bulk polarization $P_z\sim 1$. Accordingly, the corresponding component of the bulk magnetoelectric polarizability tensor is quantized,
$(\alpha_\text{ME})_{zz}=1$.

The construction presented above resembles pictorially the construction of
both one-dimensional dimerized Su-Schrieffer-Heeger model~\cite{su1979} and
Kitaev chain~\cite{kitaev2001}; Despite this similarity, we find the
obtained phase to be truly three-dimensional: the quantized surface Hall
conductance takes the same value for every orientation of the boundary as we demonstrate
numerically further below. Furthermore, the bulk magnetoelectric polarizability
tensor is found to be isotropic $\alpha_\text{ME}\equiv (\alpha_\text{ME})_{ii}=1$. This
statement is further corroborated by the existence of a truly
three-dimensional bulk topological invariant.

\textit{Bulk topological invariant $N_\mathrm{TLI}$} --- For a bulk Hamiltonian $H$,
with all eigenstates localized, the unitary $U$ that diagonalizes $H$ can be
chosen such that its matrix elements are exponentially localized in the
basis $\ket{\vR \alpha}$ 
\begin{align}
	\bra{\vR^\prime \alpha^\prime} U \ket{\vR\alpha}\sim e^{-\gamma \vert\vR-\vR^\prime\vert},
	\label{eq:exp_loc}
\end{align}
for some positive constant $\gamma$.  More concretely, we define $U$ as mapping
localized eigenstates $\ket{\psi_{n}}$ of $H$ onto atomic orbitals
$\ket{\vR\alpha}$, i.e., $U\ket{\vR\alpha}=\ket{\psi_{n(\vR,\alpha)}}$, such
that condition~(\ref{eq:exp_loc}) is satisfied. The assignment defined by
$n(\vR,\alpha)$ is not unique~\footnote{$U(1)$ gauge freedom and non-uniqueness
of $n(\vR,\alpha)$ do not affect the value of the bulk invariant.}. The bulk
integer invariant can be expressed~\footnote{We note that similar invariant appears in the anomalous Floquet-Anderson insulator~\cite{Titum2016} and its multi-drive generalization~\cite{PhysRevLett.126.106805}.} $N_\text{TLI}=\nu[U]$,
\begin{equation}
\nu[U] = \frac{i\pi\epsilon^{ijk}}{3V} \text{Tr}\left(U^\dagger[\hat X_{i},U]U^\dagger[\hat X_{j},U]U^\dagger[\hat X_{k},U]\right),
\label{eq:NTLI}
\end{equation}
with $V=N_xN_yN_z$ being the volume of the system and $\hat X_i =
\sum_{\vec{R}\alpha}R_i \ket{\vec{R}\alpha}\bra{\vec{R}\alpha}$ the $i$th
component of the position operator, $i=1,2,3$, $\vR = (x,y,z)$, and the summation over
repeated indices is assumed.  The above expression is guaranteed to take
integer values as long as~(\ref{eq:exp_loc})
is satisfied~\cite{song2014}. Note that for a finite system size, the commutators $[\hat
X_{i},U]$ need to be approximated, see~\cite{SM}. It can be analytically shown that
$N_{\text{TLI}}=1$ for the model~\eqref{eq:Hzz1}, see~\cite{SM}.

\textit{Expression for the boundary Hall conductance $\sigma_{12}^\partial$} --- We
consider a slab geometry with the width much larger than the localization
length. Let us denote the eigenstates of $H$ that are localized on one of the
two surfaces by $\{\ket{\psi^\text{surf}_n}\}$, irrespective of their energy.
Since all the bulk states are localized, some of these states can be safely
included in the set $\{\ket{\psi^\text{surf}_n}\}$ without affecting the resulting surface Chern number. The Hall conductance of the
system when the states $\{\ket{\psi^\text{surf}_n}\}$ are filled with
electrons is given by the Chern number~\cite{prodan2011}
$\sigma_{12}^\partial=\frac{e^2}{h}\text{Ch}[{\cal P}^\text{surf}]$,
\begin{equation}
\text{Ch}[\mathcal{P}] = \frac{2\pi i}{N_1N_2}\text{Tr}\left(\mathcal{P} \left[\left[\hat X_1,\mathcal{P}\right],\left[\hat X_2,\mathcal{P}\right]\right]\right),
\label{app:surfacechern}
\end{equation}
where ${\cal
P}^\text{surf}=\sum_n\ket{\psi^\text{surf}_n}\bra{\psi^\text{surf}_n}$ and
$\hat X_{1,2}$ are the two components of the position operator along the slab.
When the matrix elements of ${\cal P}^\text{surf}$ are exponentially localized,
analogous to condition~(\ref{eq:exp_loc}) with ${\cal P}^\text{surf}$ in place
of $U$, $\text{Ch}[{\cal P}^\text{surf}]$ is guaranteed to take integer values~\cite{prodan2011}.

The bulk-boundary correspondence of TLIs takes the following form
\begin{align}
	\sigma_{12}^\partial=N_\text{TLI}\frac{e^2}{h}.
	\label{eq:bb}
\end{align}
We demonstrated that the above relation holds for the model above
Eq.~(\ref{eq:Hzz1}), for the $z$-terminated crystal. Below we demonstrate
numerically that it also holds for the $x$- and $y$-terminated crystals
(hard-wall boundary), as well as for a perturbed version of the
model~(\ref{eq:Hzz1}). The general proof of  relation~(\ref{eq:bb}) for an arbitrary model of a TLI in the same phase as~(\ref{eq:Hzz1}) directly follows, as the surface Hall conductance can change only if delocalized states move to the surface, which is forbidden for TLIs in the same phase as all the states in the bulk are localized.



The quantized Hall conductance of a TLI's boundary comes together with the
quantized (isotropic) magnetoelectric polarizability coefficient
$\alpha_\text{ME}$ of its bulk. This follows directly from the arguments
presented in Ref.~\onlinecite{lapierre2021}: when the slab is fully filled with
electrons, the electrons are ``inert'' and do not respond to
an external magnetic field.  Since the boundary has non-zero quantized Hall
conductance, it follows from the Streda theorem~\cite{streda1982} that the filling of the boundary
changes by an integer amount ($\sigma_{12}^\partial h/e^2$) when the flux $\Phi_0=h/e$
threads the slab. This charge needs to be compensated by the bulk; From this
compensation, it follows that the bulk, fully filled with electrons, has an
isotropic and quantized magnetoelectric polarizability tensor
$\alpha_\text{ME}=N_\text{TLI}$. We finally note that an alternative construction of TLI can be found in App.~\ref{app:hopfconstruction}, based on the Hopf insulator~\cite{moore2008}. In particular, the notion of surface Chern number appears in the
context of Hopf insulator~\cite{PhysRevB.103.045107}, although in that context, it strongly
relies on the translation symmetry.

\textit{Numerical results} --- We perturb the model in Eq.~(\ref{eq:Hzz1}) by including nearest-neighbor hopping that
eventually push the TLI into a metallic phase. We define $H(\lambda)=H+\lambda H^\prime$ with
\begin{align}
	H^\prime&= \sum_{\langle\vec{R}\vR^\prime\rangle\alpha}\left(t_1\ket{\vec{R}^\prime\alpha}\bra{\vec{R}\alpha}+t_2\ket{\vec{R}^\prime\alpha}\bra{\vec{R}\bar\alpha}\right)  +\text{h.c.},
\label{eq:deformedhamiltonian}
\end{align}
where $\bar1 = 2$, $\bar2 = 1$, and $\langle\vR,\vR^\prime\rangle$ denotes summation over all 
nearest-neighbor pairs of lattice vectors. Below, we set $t_1=10t_2=1$, $W=1$, and consider
one-parameter family of the Hamiltonians $H(\lambda)$, $0\leq\lambda\leq1$.

\begin{figure}[t]
	\centering
	\includegraphics[width=\columnwidth]{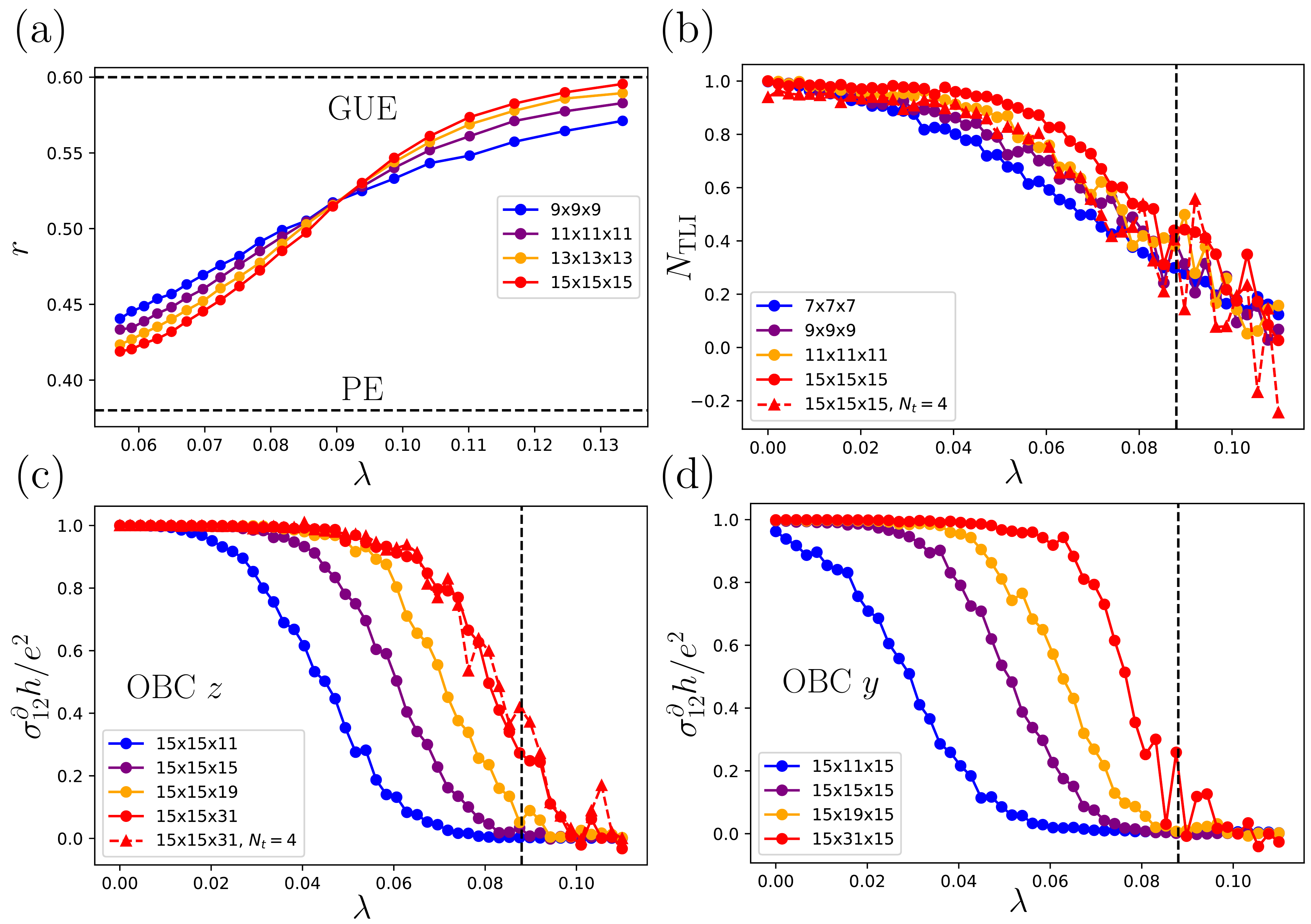}
	\caption{(a) Estimate of $\lambda_c$ from mean level spacing ratio $r$ as function of $\lambda$. $r$ is computed over 500 eigenstates in the middle of the spectrum, considering $10^3$ disorder realizations. (b) The winding number $N_{\text{TLI}}$ is computed for different system sizes with increasing hopping strength $\lambda$, where the quantized plateau increases with system size for $\lambda<\lambda_c$. (c) Surface Chern number for the hard-wall termination in $z$-direction as function of $\lambda$. (d) Same as (c), for the termination in $y$-direction. No disorder averaging was performed for the winding and Chern numbers, apart from the blue curves where average was performed over 5 disorder realizations as well as the red curves in (b) and (c) for $\lambda<\lambda_c$. The dashed curves in panels c-d are for the finite-range hopping version of the model~(\ref{eq:Hzz1}) with $N_t=4$, see the main text.}
\label{fig:numerics}
\end{figure}
We first study the metal-insulator transition using level spacing statistics.
To this end, we compute the average level spacing
ratio~\cite{oganesyan2007,suntajs2021} $r=\langle\langle
r_{n}\rangle_{n}\rangle_W$ around the middle of the spectrum, where $r_{n} =
\min\{s_{n},s_{n+1}\}/\max\{s_{n},s_{n+1}\}$ with
$s_{n}=E_{n+1}-E_{n}\geq 0$ being the level spacing and $E_{n}$, $n=1,\cdots, 2V$, the ordered eigenvalues of  $H(\lambda)$. For
$\lambda<\lambda_c$,  $r$ decreases with
increasing system size, approaching the value $r_{\text{PE}}\approx 0.38$ of the
Poisson Ensemble (PE). All eigenstates are localized in this regime. For $\lambda>\lambda_c$, $r$ increases with increasing
system size, approaching the value $r_{\text{GUE}}\approx 0.60$ of the Gaussian
Unitary Ensemble (GUE). The system is in a metallic phase
at half-filling (a mobility edge appears). From this one-parameter scaling, we find the
metal-insulator transition at half-filling occurs at $\lambda_c\approx
0.088\pm0.002$, see Fig.~\ref{fig:numerics}a.

As long as $H(\lambda)$ is in the localized phase, the
condition~(\ref{eq:exp_loc}) is satisfied and $N_\text{TLI}$ and $\sigma_{12}^\partial$
are guaranteed to take integer values for large enough system
size~\cite{prodan2011,song2014}. In practice, for the system sizes we consider
in Fig.~\ref{fig:numerics}, we find that the quantization of $N_\text{TLI}$ and
$\sigma_{12}^\partial h/e^2$ breaks before the value of $\lambda_c$ is reached. In
Fig.~\ref{fig:numerics}, we show that the range of $\lambda$ for which
$N_\text{TLI}$ and $\sigma_{12}^\partial h/e^2$ are quantized extends as the system
size is increased, showing the tendency towards the value $\lambda_c$.

We note that although the tight-binding model~\eqref{eq:Hzz1} has exponentially decalying hopping amplitudes (Fig.~\ref{fig:const}), the system remains in the same phase for a finite-range hopping version of this model. We have checked this explicitly by truncating the exponential tail of the wavefunctions $\ket{w_{\vR\alpha}}$, i.e., setting $C_{\vR'\alpha'}^{\vR\alpha}=0$ for $|\vR-\vR'|>N_t$. The results for $N_t=4$ are given by dashed curves in Fig.~\ref{fig:numerics}c--d, from where one can conclude that the phase is stable under such truncation, although the truncation increases the localization length as indicated by less good quantization.

\textit{Quantized Hall response} --- The quantized response of a TLI can be probed in the Corbino geometry by adiabatic flux insertion. Since the TLI's bulk is fully localized, the boundary can be doped with electrons independently of the bulk. The boundary of a TLI, fully filled with electrons, gives the same response to flux insertion as a torus-shaped, half-filled two-dimensional Chern insulator (extrinsic second-order TI~\cite{geier2018}), see~\cite{SM}. The main difference between these two setups is that the response of a TLI remains quantized under an arbitrarily strong boundary disorder, whereas a half-filled Chern insulator gets trivialized above certain critical disorder strength~\cite{prodan2011}.

\begin{figure}[t]
	\centering
	\includegraphics[width=\columnwidth]{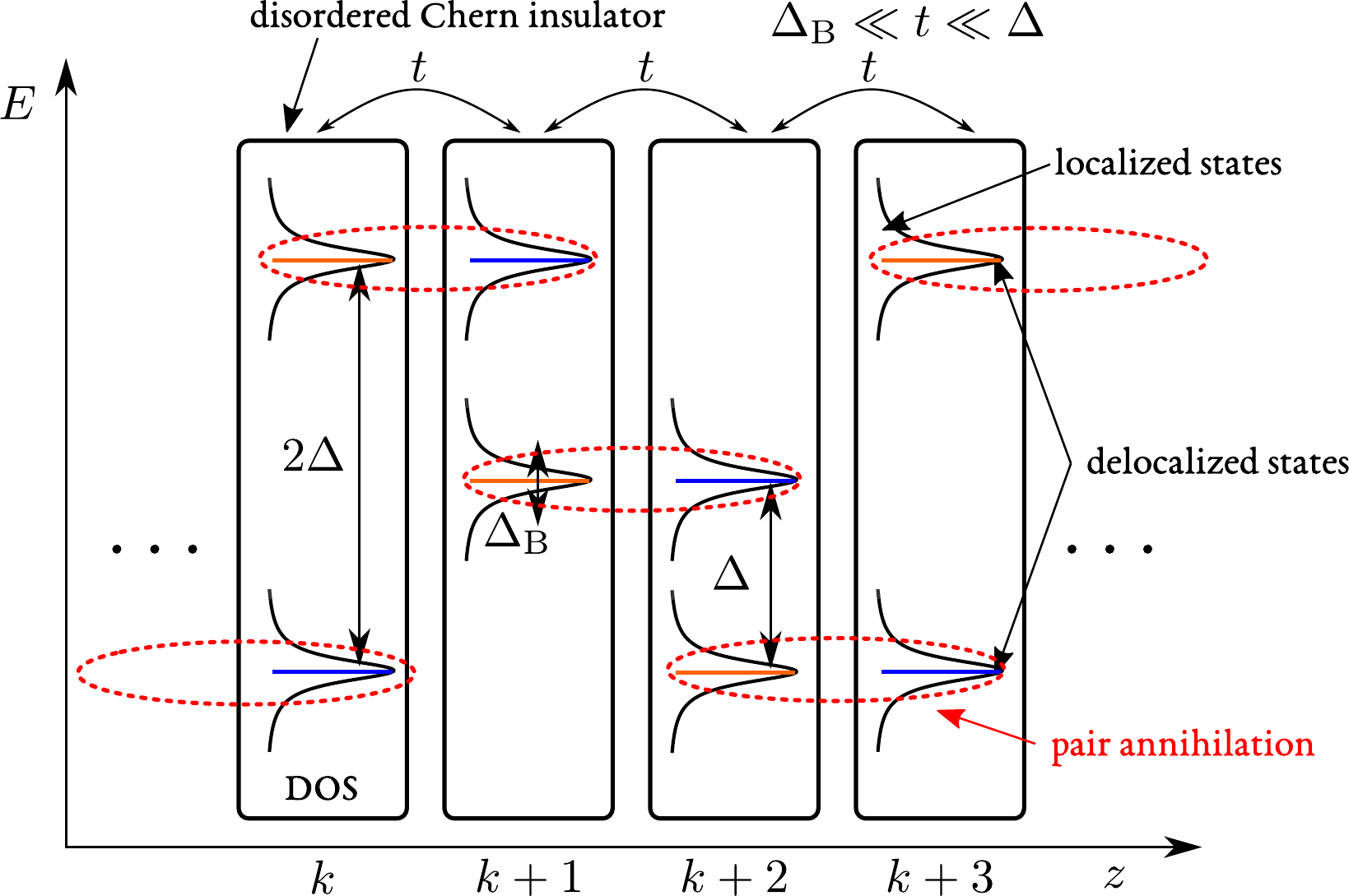}
	\caption{Sketch of the layer construction for the three-dimensional TLI. A stack along $z$-direction of two-band disordered Chern insulators, with the bandwidth $\Delta_B$ and the gap $\Delta$ or $2\Delta$, becomes fully localized due to interlayer coupling $t$ such that $\Delta_B\ll t\ll \Delta$. For each layer, density of states (DOS) is shown, and the delocalized states depicted in orange (blue) carry a positive (negative) quantized Hall conductance.}
\label{fig:TLI_model}
\end{figure}

\textit{Resonant energy model for a TLI} --- In the model defined in Eq.~\eqref{eq:Hzz1} disorder is introduced in a highly nontrivial way that is not natural to occur in experimental systems. Here we suggest an alternative model realization of a TLI can be obtained by stacking two-dimensional Chern insulators with generic disorder. We assume that each Chern insulator has two bands and a band gap $\Delta$ or $2\Delta$ (depending on the layer) that is much larger than the band width. Disorder will broaden the ``bands'' to a band with $\Delta_{\mathrm{b}}$, but should be weak enough to maintain $\Delta_{\mathrm{b}}\ll\Delta$. There is a single energy per band~\cite{prodan2011} where delocalized states appear. We mark these energies in blue or orange in Fig.~\ref{fig:TLI_model}, depending on the sign of their quantized Hall conductance. The parameters of the layers repeat with the period three, and the system is tuned such that differently colored delocalized states from neighbouring layers are on-resonance. When only resonant interlayer coupling is considered, the model is dimerized, with each dimer being a fully localized phase since it belongs to a two-dimensional unitary class and has zero Hall conductance. Hence, the delocalized states from neighbouring layers ``pair annihilate", leaving only unpaired delocalized states on the boundary of the resulting three-dimensional system. We anticipate the localization length (in the $xy$-plane) of each dimer to be rather large in this model due to exponential sensitivity of the localization length in two-dimensional unitary class~\cite{furusaki1999}. The off-resonant interlayer coupling tends to delocalize some bulk states, but we expect for strong enough disorder and for large enough flatness ratio $\Delta/\Delta_B$ (see Fig.~\ref{fig:TLI_model}) that the system is in the same phase as the above-mentioned dimerized limit.  The detailed study of the model is left for the future works.

\textit{Conclusions} --- Topology of tenfold-way (strong) TIs~\cite{schnyder2008,schnyder2009,kitaev2009} poses an obstruction to localization by disorder of its occupied bulk states~\cite{chalker1988,morimoto2015,onoda2007}. Correspondingly,
Topological Anderson Insulators~\cite{li2009} are
Anderson insulators only at certain but not all fillings
and the topological charge is carried by their delocalized bulk
states only. Our main result is finding that fully-localized insulators can be topologically non-trivial, too. In particular, we construct a three-dimensional model with broken time-reversal symmetry, where topology poses an obstruction to localization of its fully-localized bulk states all the way down to the atomic limit. We find that the three-dimensional fully-localized insulator, with broken time-reversal symmetry, does not represent a single phase of matter, but rather contains infinitely many phases that are labeled by integers. Here,
the Anderson model of localization, in the absence of mobility edge, corresponds to topologically trivial insulator and is labelled by $N_\text{TLI}=0$, whereas topologically non-trivial localized insulators are guaranteed to host states delocalized along the crystal's insulating boundary that give rise to the quantized Hall conductance~(\ref{eq:bb}) in Corbino geometry. Crucially, these delocalized boundary states remain topologically protected in the presence of an arbitrarily strong boundary disorder. We introduced the method to construct these
novel insulators, which is readily generalizable to include additional symmetries (e.g. time-reversal). We hope that this work will motivate experiments where quantized responses are observed in out-of-equilibrium settings. 

\begin{acknowledgments}\textit{Acknowledgments} --- The authors acknowledge stimulating discussions with {\"O}.~M. Aksoy, P.~W.~Brouwer, A.~Furusaki, and C.~Mudry.
We are thankful to P.~W.~Brouwer for pointing up similarities between TLIs and extrinsic higher-order TIs. BL acknowledges funding from the European Research Council (ERC) under the European Union’s Horizon 2020 research and innovation program (ERC-StG-Neupert-757867-PARATOP). LT acknowledges financial support from the FNS/SNF Ambizione Grant No. PZ00P2\_179962.
\end{acknowledgments}
\bibliographystyle{apsrev4-1}
\bibliography{ref}

\clearpage
\begin{widetext}
.
\setcounter{equation}{0}
\renewcommand\theequation{S\arabic{equation}}
\section*{Supplementary Material}

\subsection{Computation of the winding number for the model of TLI} 
\label{app:windingnumbercomputation}
In this section we show that the third winding number for the model~(1) is quantized to a non-zero value given by the Chern number of the blue subspace of a single layer (see Fig.~2a of the main text). Although the model~(1) lacks translational symmetry, its eigenstates $\ket{w_{\vec R\alpha}}$ are related to each other by translational symmetry. Below, we first prove the following relation for the unitary $U=\bigoplus_\vk U_\vk$ that diagonalizes the model~(1)
\begin{align}
	U_\vk={\cal U}_{k_xk_y}{\cal D}{\cal U}_{k_xk_y}^\dagger,
	\label{eq:Uvk}
\end{align}
where ${\cal U}_{k_xk_y}$ is a unitary that diagonalizes the Bloch Hamiltonian of
the two-dimensional, two-band Chern insulator whose positive (negative) energy states define the blue (orange) subspace in Fig.~(2a) of the main text; The matrix ${\cal
D}=\text{diag}(e^{ik_z},1)$ translates the Bloch eigenstate with the positive energy by one unit cell in $z$-direction. The Bloch states with negative energy at layer $z$, and positive energy at layer $z+1$ are
\begin{equation}
\begin{cases}
\ket{u^-_{zk_xk_y}}=a_{k_xk_y}\ket{z1}+b_{k_xk_y}\ket{z2},\\
\ket{u^+_{(z+1)k_xk_y}}=b_{k_xk_y}^*\ket{(z+1)1}-a_{k_xk_y}^*\ket{(z+1)2},
\end{cases}
\end{equation}
where we used the notation $\mathcal{U}_{k_xk_y}=\begin{pmatrix} a_{k_xk_y}&b_{k_xk_y}^*\\
b_{k_xk_y}&-a_{k_xk_y}^* \end{pmatrix}$. The two states above are discontinuous over the BZ. We consider linear combination of the two states that is continuous
\begin{equation}
\begin{cases}
\ket{\tilde{u}^-_{k_xk_y}}=a_{k_xk_y}^*\ket{u^-_{zk_xk_y}}+b_{k_xk_y}\ket{u^+_{(z+1)k_xk_y}},\\
\ket{\tilde{u}^+_{k_xk_y}}=b^*_{k_xk_y}\ket{u^-_{zk_xk_y}}-a_{k_xk_y}\ket{u^-_{(z+1)k_xk_y}}.
\end{cases}
\end{equation}
After performing the Fourier transform in the $z$-direction, we conclude that the continuous unitary is of the form \eqref{eq:Uvk}. In order to compute the third winding of the above unitary, we use Eq.~(\ref{app:3rdwindingnumber}) which gives eight different terms. We can group four of them as
\begin{equation}
2i \text{Tr}\left(\text{diag}(1,0)\left[\mathcal{U}_{k_xk_y}^{\dagger}(\partial_{k_x}\mathcal{U}_{k_xk_y}),\mathcal{U}_{k_xk_y}^{\dagger}(\partial_{k_y}\mathcal{U}_{k_xk_y})\right]\right) = 2i \left[\mathcal{U}_{k_xk_y}^{\dagger}(\partial_{k_x}\mathcal{U}_{k_xk_y}),\mathcal{U}_{k_xk_y}^{\dagger}(\partial_{k_y}\mathcal{U}_{k_xk_y})\right]_{00},
\label{app:term1}
\end{equation}
additional two terms as
\begin{equation}
ie^{-ik_z}\left((\partial_{k_x}\mathcal{U}^{\dagger}_{k_xk_y})\mathcal{U}_{k_xk_y}\mathcal{D}\mathcal{U}_{k_xk_y}^{\dagger}(\partial_{k_y}\mathcal{U}_{k_xk_y})-\mathcal{U}_{k_xk_y}^{\dagger}(\partial_{k_y}\mathcal{U}_{k_xk_y})\mathcal{D}(\partial_{k_x}\mathcal{U}_{k_xk_y}^{\dagger})\mathcal{U}_{k_xk_y})\right)_{00},
\label{app:term2}
\end{equation}
and the two last terms are
\begin{equation}
ie^{ik_z}\left((\partial_{k_x}\mathcal{U}_{k_xk_y}^{\dagger})\mathcal{U}_{k_xk_y}\mathcal{D}^{*}\mathcal{U}_{k_xk_y}^{\dagger}(\partial_{k_y}\mathcal{U}_{k_xk_y})-\mathcal{U}_{k_xk_y}^{\dagger}(\partial_{k_y}\mathcal{U}_{k_xk_y})\mathcal{D}^{*}(\partial_{k_x}\mathcal{U}_{k_xk_y}^{\dagger})\mathcal{U}_{k_xk_y})\right)_{00}.
\label{app:term3}
\end{equation}
The terms~(\ref{app:term1})-(\ref{app:term2}) cancel out after integrating over $k_z$ from 0 to $2\pi$. Hence, we consider only the terms in Eq.~\eqref{app:term1}. Using the fact that the Chern number for the lower band reads
\begin{equation}
\text{Ch} = \int_{\text{BZ}} \frac{\text{d}^{2}k}{2\pi i} \left(\bra{\partial_{k_x}u^{-}_{k_x,k_y}}\partial_{k_y}\ket{u^{-}_{k_x,k_y}} -\bra{\partial_{k_y}u^{-}_{k_x,k_y}}\partial_{k_x}\ket{u^{-}_{k_x,k_y}} \right)
\end{equation}
together with $\mathcal{U}_{k_xk_y}\ket{0} =\ket{u^{-}_{k_xk_y}}$, and  $\mathcal{U}_{k_xk_y}^{\dagger}(\partial_{k_x}\mathcal{U}_{k_x,k_y})\mathcal{U}_{k_xk_y}^{\dagger}(\partial_{k_y}\mathcal{U}_{k_xk_y}) = -(\partial_{k_x}\mathcal{U}_{k_xk_y}^{\dagger})(\partial_{k_y}\mathcal{U}_{k_xk_y})$, we obtain 
\begin{equation}
\nu[U_{\vk}] = \text{Ch}.  
\end{equation}
We note that in the case where $\mathcal{D} = \text{diag}(e^{i l k_z},1)$ with $l\in\mathbb{Z}$, the obtained relation becomes $\nu[U_{\vk}] = l\text{Ch}$.

\subsection{Third winding and Chern numbers} 
\label{app:windingnumberconstruction}

For translationally invariant unitaries $U=\bigoplus_\vk U_\vk$, in the thermodynamic limit, the third winding number is defined  
\begin{equation}
\nu[U_{\vk}] = \int_{\text{BZ}}\frac{\text{d}^3k}{8\pi^2}\text{Tr}\left(U_{\vk}^{\dagger}\partial_{k_x}U_{\vk}\left[U_{\vk}^{\dagger}\partial_{k_y}U_{\vk},U_{\vk}^{\dagger}\partial_{k_z}U_{\vk}\right]\right),
\label{app:3rdwindingnumber}
\end{equation}
while for finite systems, the integral is replaced by a sum over discrete momenta $\vk$. For unitaries that lack translational symmetry but satisfy condition~(4), the third winding number can be defined~\cite{song2014}
\begin{equation}
\nu[U] = \frac{i\pi}{3}\frac{1}{N_xN_yN_z}\epsilon^{ijk} \text{Tr}\left(U^{-1}[\hat X_{i},U]U^{-1}[\hat X_{j},U]U^{-1}[\hat X_{k},U]\right),
\label{app:windingnumber}
\end{equation}
where $\hat X_i = \sum_{\vec{R}\alpha}R_i \ket{\vec{R}\alpha}\bra{\vec{R}\alpha}$ is the position operator in direction $i=1,2,3$, $\vR=(x,y,z)$, and summation over repeated indices is assumed.  The above expression takes non-zero integer values only in the thermodynamic limit. In order to apply it to finite systems, we approximate the commutators $[X_i,U]$ by $\lfloor X_i,U\rfloor$,  i.e., linear combination of unitaries with integer number of flux quanta inserted
\begin{equation}
\lfloor \hat X_i,U\rfloor = \sum_{m=1}^{N_i-1}c_m e^{2\pi i m \hat X_i/N_i}Ue^{-2\pi i m \hat X_i/N_i}.
\label{app:braket}
\end{equation}
The choice of coefficients $c_m$, giving the correct thermodynamic-limit value of $[X_i,U]$, is not unique~\cite{prodan2011, song2014}. 
In general, these are the coefficients of the discrete Fourier transform of a function $f(x)$ defined on the circle $[-\floor{\frac{N_i}{2}},\floor{\frac{N_i}{2}}]$ that is periodic and equal to $f(x)=x$ for $x<\alpha \lesssim\floor{\frac{N_i}{2}}$.
In this work, we take $c_m$ to be the Fourier coefficients of the function $f(x)=x$,
\begin{equation}
c_m = \frac{e^{2\pi i m (\floor{\frac{N_i}{2}}+1) /N_i}}{1-e^{2\pi i m /N_i}}.
\end{equation}

Similarly, for a translationally invariant projector ${\cal P}=\bigoplus_{k_xk_y} {\cal P}_{k_x k_y}$, in the thermodynamic limit, the Chern number of the Hilbert space, onto which this projector projects, is defined~\cite{ryu2010}
\begin{equation}
\text{Ch}[\mathcal{P}_{k_x k_y}]= i\int_{\text{BZ}} \frac{\text{d}k_x\text{d}k_y}{2\pi}\text{Tr}\left(\mathcal{P}_{k_x k_y}\left[\partial_{k_x}\mathcal{P}_{k_x k_y},\partial_{k_y}\mathcal{P}_{k_x k_y}\right]\right),
\end{equation}
for finite systems, the above integral is replaced by a sum over discrete momenta $(k_x,k_y)$. When the projector $\mathcal{P}$ lacks translational symmetry, but satisfied the condition
\begin{align}
	\bra{\vR^\prime \alpha^\prime} \mathcal{P} \ket{\vR\alpha}\sim e^{-\gamma \vert\vR-\vR^\prime\vert},
	\label{eq:exp_locP}
\end{align}
for some positive $\gamma$, the Chern number in $xy$ plane in the therodynamic limit can be defined~\cite{prodan2011}
\begin{equation}
\text{Ch}[\mathcal{P}] = \frac{2\pi i}{N_xN_y}\text{Tr}\left(\mathcal{P} \left[\left[\hat X_1,\mathcal{P}\right],\left[\hat X_2,\mathcal{P}\right]\right]\right).
\label{app:surfacechern}
\end{equation}
For finite systems, the commutator $\left[\hat X_1,\mathcal{P}\right]$ is approximated by $\lfloor \hat X_i,P\rfloor$, see Eq.~(\ref{app:braket}).

\subsection{Details on numerical calculations}
For the numerical calculations in this work, the projectors $\mathcal{P}_z^\pm$ used for the model~(1) are obtained as the projectors onto one of the two bands of the following model for Chern insulator
\begin{equation}
H_{2D}= \sin(k_x)\sigma_x + \sin(k_y)\sigma_y + (1-\cos(k_x)-\cos(k_y))\sigma_z.
\label{2dhamiltonian}
\end{equation}
As explained in the main text, the eigenstates of the Hamiltonian~(1), are
\begin{align}
    \ket{w_{\vec{R}\alpha}}=(\mathcal{P}_{z}^-+\mathcal{P}_{z+1}^+)(\ket{\vec{R}\alpha}+\ket{(\vec{R}+\hat{e}_z)\alpha}),
\label{app:wannierfunctions}
\end{align}
with eigenenergy $W_{\vR\alpha}$. Hence,
\begin{equation}
H = \sum_{\vec{R}\alpha}W_{\vec{R},\alpha}\ket{w_{\vec{R}\alpha}}\bra{w_{\vec{R}\alpha}}
\label{app:hamitlonianmodel2}
\end{equation}
In other words, the above Hamiltonian $H = UH_WU^{\dagger}$ is unitarily related to Hamiltonian $H_W$ of a trivial localized insulator, with the unitary $U$ satisfying the localization condition~(4).
To show the orthonormality relation $\langle  w_{\vR'\alpha'}|w_{\vR\alpha}\rangle = \delta_{\vR\vec{R}'}\delta_{\alpha\alpha'}$, we denote $\vR = (\vr,z)$, with $\vr$ two-dimensional lattice vector, and consider the scalar product
\begin{equation}
\begin{split}
\langle w_{\vr'z\alpha'}|w_{\vr z\alpha}\rangle &= \left(\bra{\vr' z \alpha'}+\bra{\vr' (z+1) \alpha'}\right)(\mathcal{P}_{z}^++\mathcal{P}_{z+1}^-)\left(\ket{\vr z \alpha}+\ket{\vr (z+1) \alpha}\right)\\
&= \bra{\vr' z \alpha'}\mathcal{P}_{z}^+\ket{\vr z \alpha} + \bra{\vr' z \alpha'}\mathcal{P}_{z}^-\ket{\vr z \alpha} = \delta_{\vec{r}\vec{r}'}\delta_{\alpha\alpha'},
\end{split}
\end{equation}
where we used that $\mathcal{P}_z^+ + \mathcal{P}_z^- = \mathbb{I}$ on the subspace of the Hilbert space corresponding to the layer $z$. For $z\neq z'$, it is sufficient to consider $z' = z+1$,
\begin{equation}
\begin{split}
\langle w_{\vr (z+1)\alpha}|w_{\vr z\alpha}\rangle  &= (\bra{\vr (z+1)\alpha}+\bra{\vr (z+2)\alpha})(\mathcal{P}^+_{z+1}+\mathcal{P}^-_{z+2})(\mathcal{P}^+_{z}+\mathcal{P}^-_{z+1})(\ket{\vr z\alpha}+\ket{\vr (z+1)\alpha})\\
&= \bra{\vr (z+1)\alpha}\mathcal{P}^+_{z+1}\mathcal{P}^-_{z+1}\ket{\vr (z+1)\alpha}=0.
\end{split}
\end{equation}
In the above derivation, we used that  $\mathcal{P}^+_{z+1}\mathcal{P}^-_{z+1}=0$.

Below, we explain how the unitary $U_W$ of the perturbed model $H(\lambda)$ is constructed
\begin{equation}
H(\lambda) =  \sum_{\vec{R}\alpha}\left[W_{\vec{R},\alpha}\ket{w_{\vec{R}\alpha}}\bra{w_{\vec{R}\alpha}} + \lambda\sum_{i=1}^3 \left(t_1\left[\ket{\vec{R}+\hat{e}_i,\alpha}\bra{\vec{R},\alpha}+\text{h.c.}\right] +t_2\left[ \ket{\vec{R}+\hat{e}_i,\alpha}\bra{\vec{R},\bar\alpha}  +\text{h.c.}\right]\right) \right],
\label{app:deformedhamiltonian}
\end{equation}
where we set $t_1=10 t_2=1$. All $2V$ eigenstates $\{\ket{\psi_n}\}$ of $H(\lambda)$  are localized for $\lambda < \lambda_c$ in the thermodynamic limit. The unitary $U_W$ is constructed from these localized eigenstates, after sorting them such that $U_W$ maps $\ket{\vec{R}\alpha}$ to the two eigenstates $\ket{\psi_{n(\vec{R},\alpha)}}$, localized closest to the orbitals at $\ket{\vec{R}\alpha}$. To this end, we compute $\text{arg}(\bra{\psi_n}e^{2\pi i \hat X_j/N_j}\ket{\psi_n})$, where the operators $e^{2\pi i \hat X_j/N_j}$ are compatible with PBC. This way, we obtain the following coordinate vectors on the 3-torus ($\mathbb{T}^3$)
\begin{equation}
\vec{v}(\ket{\psi_n}) = \left(\frac{N_x}{2\pi}\text{arg}(\bra{\psi_n}e^{2\pi i \hat X/N_x}\ket{\psi_n}),\frac{N_y}{2\pi}\text{arg}(\bra{\psi_n}e^{2\pi i \hat Y/N_y}\ket{\psi_n}),\frac{N_z}{2\pi}\text{arg}(\bra{\psi_n}e^{2\pi i \hat Z/N_z}\ket{\psi_n})\right)^T,
\label{coordinatevectortorus:app}
\end{equation}
where we choose the branch cut of $\text{arg}(\xi)$ to be $[0,\infty)$, such that $0\leq \text{arg}(\xi)<2\pi$.
The sorting procedure is the following: we first minimize the distance between $\vv(\ket{\psi_n})$ and $\vec{v}(\ket{\vR\alpha})$ by computing
\begin{equation}
||\vec{v}(\ket{\vR\alpha})-\vec{v}(\ket{\psi_n})||_{\mathbb{T}^3}=\sqrt{\sum_{i=1}^3d\left(R_i, (\vec v(\ket{\psi_n})_i\right)^2}
\label{app:metic}
\end{equation}
where $d(x,y)$ is the following metric on the circle of radius $\frac{N_i}{2\pi}$,
\begin{equation}
d(x,y) = \text{min}\{|x-y|,N_i-|x-y|\}.
\end{equation}
To each orbital $\ket{\vec{R}\alpha}$, we associate eigenvector $\ket{\psi_n}$ closest to it according to the metric~(\ref{app:metic}), unless this eigenvector is already assigned to a different lattice vector, in which case next-closest unassigned eigenvector is used. This defines a unitary $U_W$, $U_W \ket{\vec{R}\alpha}=\ket{\psi_{n(\vR,\alpha)}}$, satisfying the localization condition~(4). As the metallic phase is approached, the eigenstates $\ket{\psi_n}$ become plane-wave-like, implying that $|\bra{\psi_n}e^{2\pi i \hat X_j/N_j}\ket{\psi_n}|\rightarrow0$. Since $\text{arg}(\xi)$ is not well-defined at $\xi=0$, the above procedure is not well-define in a metallic phase. For finite systems, we compute $\langle\vert\vert\vv(\ket{\psi_n})\vert\vert_{\mathbb{T}^3}\rangle_n$, in order to define the region in parameter space in which the winding number is still well-defined. Such region exists for certain values of $\lambda$ which are smaller than $\lambda_c$, see Fig.~\ref{fig:norm}a.
\begin{figure}[hbt]
\begin{center}
	\includegraphics[width=15cm]{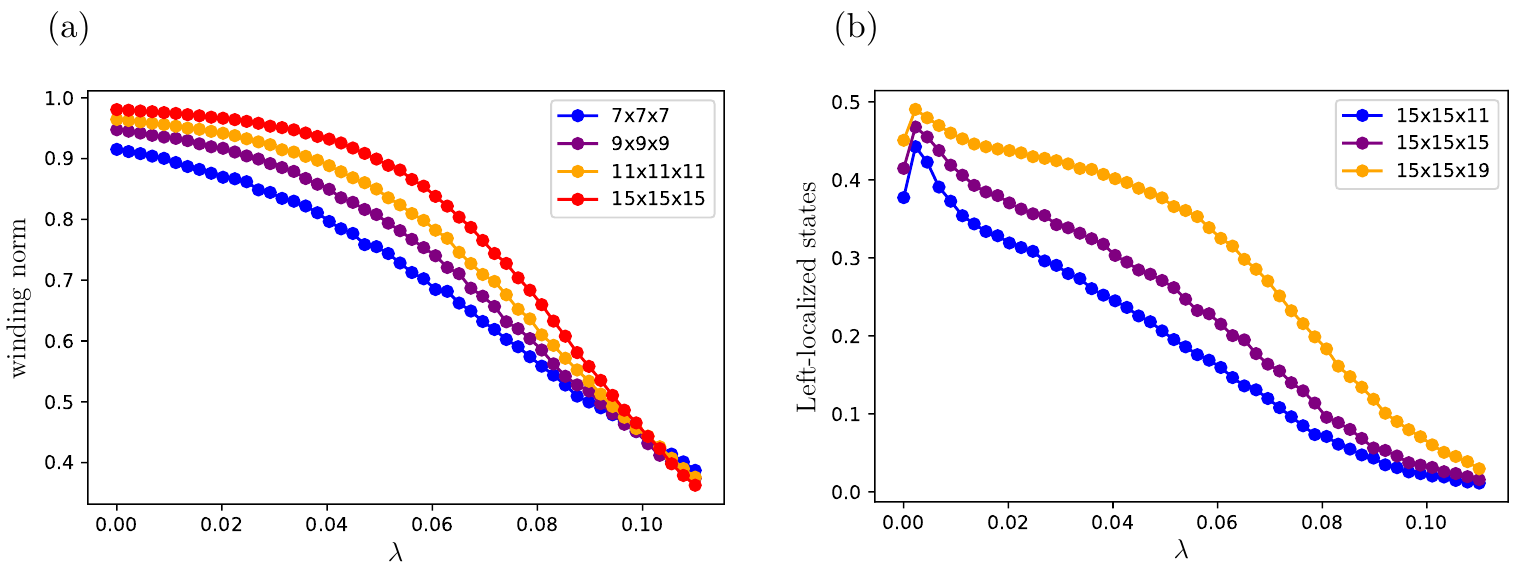}
	\caption{\label{fig:norm} (a) The average $\langle\vert\vert\vv(\ket{\psi_n})\vert\vert_{\mathbb{T}^3}\rangle_n$ over all eigenstates as function of parameter $\lambda$, see Eq.~(\ref{app:deformedhamiltonian}). This quantity goes from $1$, deep in the localized phase, to 0 in the metallic phase. (b) Portion of eigenstates that are localized in the left-half of the system, i.e. for $z\in \{1,...,\floor{N_z/2}\}$. This fraction goes from approximately one-half, at $\lambda = 0$, to $0$, for $\lambda>\lambda_c$. }
\end{center}
\end{figure}

We now define the projector $\mathcal{P}_W^{\text{surf}}$ onto the surface perpendicular to $z$-axis. This is achieved by removing localized eigenstates $\ket{\psi_n}$ within the bulk of the system, such that the obtained projector projects onto two decoupled surfaces. We first define the set of localized eigenstates in the bulk of the slab as
\begin{equation}
\text{bulk}(W) = \left\{\ket{\psi_n} ,1-\quad \sum_{\substack{x,y,\alpha,\\ z=\floor{N_z/2}-L}}^{z=\floor{N_z/2}+L}|\langle{\vec{R}\alpha}|\psi_n\rangle|^2<\epsilon \right\},
\end{equation}
where $L$ is some bulk cut-off that needs to be large enough such that the upper and lower boundaries are fully decoupled, and $\epsilon$ is taken to be $10^{-1}$. The eigenstates that do not belong to $\text{bulk}(W)$ are assigned to one of the two boundary surfaces which defines the projector $\mathcal{P}^{\text{surf}}_W$
\begin{equation}
\mathcal{P}_W^{\text{surf}}(\vec{x},\vec{x}') = \left[\delta_{\vec{x},\vec{x}'}-\sum_{\ket{\psi_n}\in\text{bulk}(W)}\psi_n(\vec{x}')^*\psi_n(\vec{x})\right]\theta\left(z-\floor{N_z/2}\right)\theta\left(z'-\floor{N_z/2}\right),
\end{equation}
where $\theta(z)$ is the Heaviside theta function, and we assumed that the plane $z=\floor{N_z/2}$ passes through the middle of the slab. As the system approaches metallic phase, portion of surface-localized states decreases to zero, see Fig.~\ref{fig:norm}b.

\subsection{Quantized Hall reponse of TLIs}
\label{app:quantizedresponse}

In this Appendix, we give further details on the measurement setup for the quantized response probing the bulk topology of TLIs using Corbino geometry. As stated in the main text, a non-zero quantized Hall conductance can be measured in genus-one (Corbino) geometry. Below we propose a setup, inspired by the setup of Ref.~\onlinecite{kundu2020}, that shows unique quantized signature of the bulk topology of TLIs.
\begin{figure}[htbp]
	\centering
	\includegraphics[width=0.7\columnwidth]{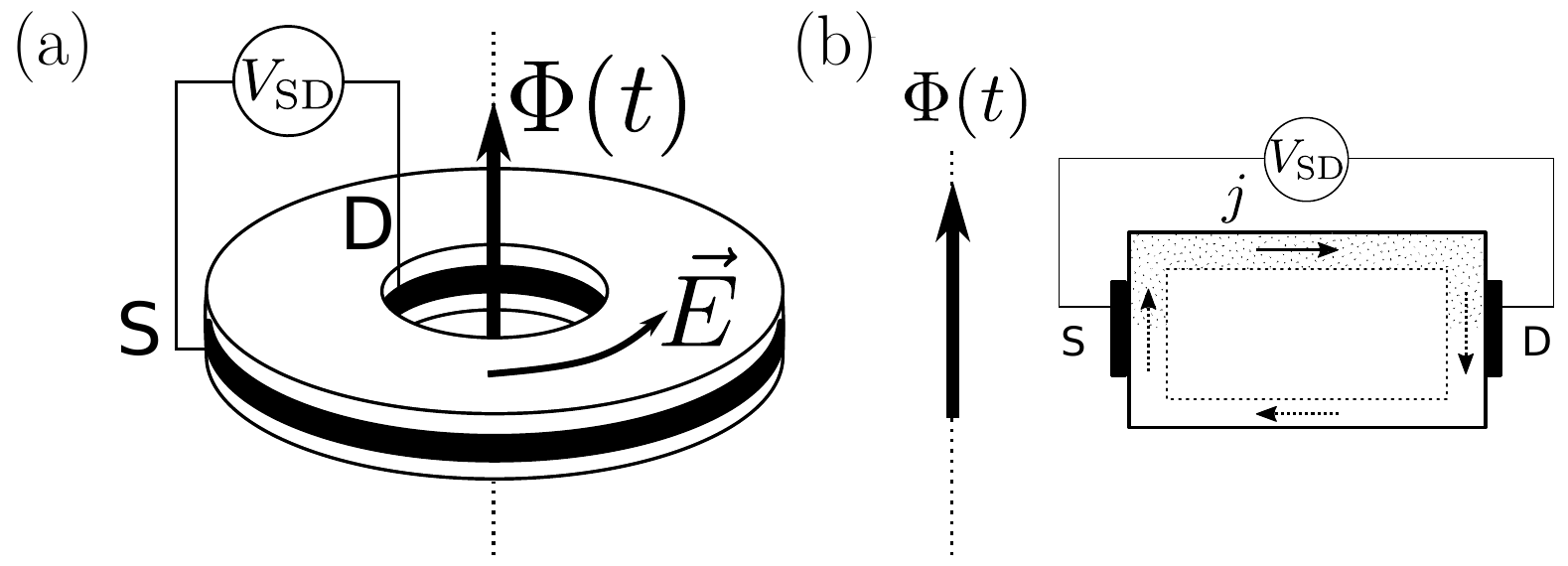}
	\caption{(a) A non-trivial TLI in Corbino geometry is threaded by
		magnetic flux $\Phi(t)=\Phi_0t/T$, with a large
		source-drain voltage $V_\text{SD}$ applied. (b) The cross section of the
		Corbino setup. As the magnetic flux is varied, the charge
		$\sigma_{12}^\partial h/e$ is being pumped via the top surface (Laughlin pump). The dotted region (top surface) is filled with electrons in the steady-state, while the bulk is assumed to be empty.}
	\label{fig:Corbino}
\end{figure}

Consider a thick Corbino disc, as shown in Fig.~\ref{fig:Corbino}a, and assume that the boundary is doped with electrons, such that \emph{all} boundary states are occupied. In this setup, an extrinsic second-order TI~\cite{geier2018}, which consists of a trivial fully-localized bulk with a Chern insulator covering (fully or partially) its boundary, does not react to (adiabatic) time-dependent magnetic flux insertion $\Phi(t)=\Phi_0t/T$. On the other hand, according to Laughlin argument, the boundary of TLIs pumps $\sigma_{12}^\partial h/e^2$ electrons during the period $T$ around its cross-section, see Fig.~\ref{fig:Corbino}b. In order to measure this charge pumping, a source-drain voltage, much larger than the bulk mobility gap, is applied. In the presence of such voltage, the charge is removed from the bottom surface during the pumping. Hence, we expect that in the steady state the top surface remains fully occupied, while the bottom one is empty, see Fig.~\ref{fig:Corbino}b. (We neglect the effects of the charging energy.) Although such a non-uniform boundary Fermi energy $E^\partial_F$ gives rise to chiral states, these states do not connect source to drain: the quantized transfer of $\sigma_{12}^\partial h/e^2$ electrons during one period $T$ occurs via pumping of the boundary states from source to drain via the top surface.

\subsection{Relation between TLIs and $N$-band Hopf insulators}
\label{app:hopfconstruction}
The models of TLIs, introduced in Eq.~(1), as well as its $N$-band generalization discussed in the main text and App.~A, become translationally invariant if we change $W_{\vR\alpha}$ from being random numbers to $W_{\vR\alpha}=\alpha$, with $\alpha=1,\dots,N$. Such deformation of the model does not change the unitary~\eqref{eq:Uvk}; The resulting, translationally invariant model, corresponds to topologically non-trivial $N$-band Hopf insulator~\cite{lapierre2021}. Nevertheless, TLIs and $N$-band Hopf insulators are distinct phases of matter: the constraint of fully-localized bulk implies that TLIs are insulating at an arbitrary filling, whereas $N$-band Hopf insulators are insulating at integer fillings while being conducting at fractional fillings. This distinction is important, since it implies that the boundary Fermi energy $E_F^\partial$ of $N$-band Hopf cannot be varied independently of that of the bulk. As a consequence, its quantized boundary response can be measured only in a transient state~\cite{lapierre2021}.

Topology of $N$-band Hopf insulator crucially depends on the presence of translational symmetry. For example, inclusion of perturbations that reduce the translational symmetry down to one of its subgroups can trivialize $N$-band Hopf insulators~\cite{lapierre2021}. Nevertheless, we show that every $N$-band Hopf insulator can be deformed to a TLI phase with the same bulk topological invariant. We illustrate this point using $N=2$ Moore-Ran-Wen model~\cite{moore2008}. The two-band Bloch Hamiltonian is defined as
\begin{equation}
h_{k_xk_yk_z}=\vec{v}\cdot\vec{\sigma},
\label{mrwmodel}
\end{equation}
with $v_i=\vec{z}^{\dagger}\sigma_i\vec{z}$, where $\vec{z}=(z_1,z_2)^T$, with $z_1=\sin(k_x)+i\sin(k_y)$ and $z_2=\sin(k_z)+i[\cos(k_x)+\cos(k_y)+\cos(k_z)-\frac{3}{2}]$. The unitary $U_\vk$, that diagonalizes above Bloch Hamiltonian, has a non-zero third winding number $\nu[U_\vk]=1$. The two normalized eigenvectors of the Bloch Hamiltonian \eqref{mrwmodel} are
\begin{align}
    \ket{u_{\vec{k}1}}&=\vert\vec{z}\vert^{-1}(z_1,z_2)^T,\nonumber\\
    \ket{u_{\vec{k}2}}&=\vert\vec{z}\vert^{-1}(z_2^*,-z_1^*)^T,
\end{align}
which are continuous functions of $\vec{k}$. We extend these two Bloch eigenvectors to the whole lattice by defining $\psi_{\vec{k}\alpha}(\vx)=e^{-i\vec{k}\cdot\vx}u_{\vec{k}\alpha}(\vx)$. Using these two Bloch eigenvector, we define Wannier functions $\ket{w_{(0,0,0)\alpha}}$, $\alpha=1,2$,
\begin{align}
	\vert w_{\vR \alpha}\rangle=\frac{1}{\sqrt{N_xN_yN_z}}\sum_\vk e^{i\vk\cdot\vR}\vert\psi_{\vk \alpha}\rangle.
	\label{eq:wRn}
\end{align}
We use the above Wannier functions to define the following disordered Hamiltonian
\begin{equation}
H_1 = \sum_{\vec{R}\alpha}W_{\vec{R}\alpha}\ket{w_{\vec{R}\alpha}}\bra{w_{\vec{R}\alpha}}.
\end{equation}
As the bulk Wannier functions of the Hopf insulator are exponentially localized, the unitary $U\ket{\vec{R}\alpha}=\ket{w_{\vec{R}\alpha}}$ satisfies localization condition~(4) and the winding number can be computed using \eqref{app:windingnumber}, $\nu[U]=\nu[U_\vk]=1$. By virtue of the bulk-boundary correspondence of the Hopf insulator~\cite{lapierre2021,PhysRevB.103.045107}, imposing open boundary conditions in any of the three spatial directions results in surface Chern number equal to $\nu[U]=1$.
\end{widetext}

\end{document}